\documentclass[aps,pra,reprint,a4paper,showpacs,superscriptaddress,floatfix]{revtex4-1}
\usepackage[pdftex]{graphicx}
\usepackage{dcolumn}
\usepackage{bm}
\usepackage[usenames, dvipsnames]{color}
\usepackage{comment}
\usepackage[at]{easylist}
\ListProperties(Progressive*=.5cm,Start1=1)

\usepackage[T1]{fontenc}

\usepackage{enumerate}
\usepackage{amsfonts}
\usepackage{amssymb}

\usepackage{amsthm}
\usepackage{bbm}
\usepackage{multirow}
\usepackage[scr=boondoxo,frak=boondox,scrscaled=1.05]{mathalfa}
\usepackage{amsmath}

\DeclareFontFamily{T1}{pzc}{}
\DeclareFontShape{T1}{pzc}{m}{it}{<-> [1.18] pzcmi8t}{}
\DeclareMathAlphabet{\mathpzc}{T1}{pzc}{m}{it}

\usepackage{hyperref}
\hypersetup{
     colorlinks   = true,
     citecolor    = blue,
     linkcolor=black,
     urlcolor=black}

\usepackage{subfigure}

\usepackage{letltxmacro}
\makeatletter
\let\oldr@@t\r@@t
\def\r@@t#1#2{%
\setbox0=\hbox{$\oldr@@t#1{#2\,}$}\dimen0=\ht0
\advance\dimen0-0.2\ht0
\setbox2=\hbox{\vrule height\ht0 depth -\dimen0}%
{\box0\lower0.4pt\box2}}
\LetLtxMacro{\oldsqrt}{\sqrt}
\renewcommand*{\sqrt}[2][\ ]{\oldsqrt[#1]{#2}}
\makeatother

\usepackage{array}
\newcolumntype{L}[1]{>{\raggedright\let\newline\\\arraybackslash\hspace{0pt}}m{#1}}
\newcolumntype{C}[1]{>{\centering\let\newline\\\arraybackslash\hspace{0pt}}m{#1}}
\newcolumntype{R}[1]{>{\raggedleft\let\newline\\\arraybackslash\hspace{0pt}}m{#1}}
\usepackage{tabularx}

\def\KeyWord#1{$\backslash$\IfColor{$\!\!$\textRed{#1}\textBlack}{#1}$\!\!$}

\usepackage{wasysym}

\newcommand{\be}{\begin{equation} }
\newcommand{\ee}{\end{equation} }
\newcommand{\ba}{\begin{eqnarray} }
\newcommand{\ea}{\end{eqnarray} }
\newcommand{\bit}{\begin{itemize}}
\newcommand{\eit}{\end{itemize}}
\newcommand{\ben}{\begin{enumerate}}
\newcommand{\een}{\end{enumerate}}

\newcommand{\bn}{\begin{easylist} 
\NewList(Indent1=.5cm,Start1=1,Hang=true,Progressive*=.5cm)
}
\newcommand{\en}{\end{easylist} }

\newcommand{\bl}{\begin{easylist}
\NewList(Indent1=.5cm,Hide=100,Style*=\scshape$\bullet\,\,$,Hang=true,Progressive*=.5cm)
}
\newcommand{\el}{\end{easylist} }

\newcommand{\x}{\sigma_{1}}

\newcommand{\z}{\sigma_{3}}
\newcommand{\tx}{\tau_{1}}

\newcommand{\tz}{\tau_{3}}

\newcommand{\tp}{\tau_{+}}
\newcommand{\tn}{\tau_{0}}

\newcommand{\cH}{\mathcal{H}}
\newcommand{\cF}{\mathcal{F}}

\renewcommand\vec{\boldsymbol}
\renewcommand{\i}{\mathrm{i}}

\newcommand{\e}{\mathrm{e}}

\def\Im{\mathrm{Im}}
\def\Re{\mathrm{Re}}

\def\E{{\vec{E}}}
\def\r{{\vec{r}}}
\def\k{{\vec{k}}}
\def\q{{\vec{q}}}
\def\p{{\vec{p}}}

\def\J{{\vec{j}}}
\def\K{{\vec{K}}}
\def\Jsc{{\hJ_\mathrm{sc}}}
\def\hJ{\vec{j}}
\def\hK{\vec{K}}
\def\hR{\vec{R}}

\def\v{{\vec{v}}}
\def\u{{\vec{u}}}

\def\bxi{\bar{\xi}}

\def\hH{H}
\def\tH{\tilde{H}}
\def\tK{\tilde{\vec{K}}}
\def\tR{\tilde{\vec{R}}}
\def\tJ{\tilde{\vec{j}}}
\def\tN{\tilde{N}}

\def\bra#1{\langle#1|}
\def\ket#1{|#1\rangle}

\def\cexp#1{\langle#1\rangle}

\def\up{\uparrow}
\def\dn{\downarrow}

\begin{document}
\title{
Supercurrent induced resonant optical response
}

\author{Philip J. D. Crowley}
\author{Liang Fu}
\affiliation{Department of Physics, Massachusetts Institute of Technology, Cambridge, Massachusetts 02139, USA}

\date{\today}

\begin{abstract}
The optical conductivity encodes the current response to a time dependent electric field. We develop a theory of the optical conductivity $\sigma(\omega)$ in presence of a dc supercurrent. Current induced optical response is prohibited by Galilean invariance from occurring in systems with a single parabolic band. However, we show that lattice effects give rise to a pronounced current dependent peak in $\sigma(\omega)$ at the gap edge $\omega = 2 \Delta$, which diverges in the clean limit. We demonstrate this in a model of a multi-band superconductor. Our theory explains the recent observation of a current induced peak in the optical conductivity in NbN by~\textcite{nakamura2019infrared}, and provides a new mechanism for direct activation of the Higgs mode with light.
\end{abstract}

\maketitle

The nonlinear electromagnetic response of quantum materials at terahertz (THz) frequencies and below allows direct probing of electronic structure~\cite{hafez2016intense}, and provides new inroads into the ``THz gap''---a frequency range that has been barely exploited for practical applications in signal generation or detection~\cite{williams2005filling}. In particular, in metals, the nonlinear Hall effect~\cite{sodemann2015quantum,ma2019observation,du2021nonlinear} and the linear/circular photogalvanic effects~\cite{moore2010confinement,deyo2009semiclassical,de2017quantized,ji2019spatially,golub2020semiclassical} 
at low frequency provide direct measurements of Fermi surface properties without generating high-energy excitations, and have established new connections between nonlinear conductivity and the quantum geometry of Bloch electrons~\cite{tokura2018nonreciprocal,ma2021topology,orenstein2021topology}.

In the context of superconductors, the nonlinear response has long been a focus of interest~\cite{gorkov1969behavior,gorkov1969superconducting,amato1976measurement,gor1996generalization,gradhand2013kerr,silaev2019nonlinear,xu2019nonlinear}, where effects second order in an ac electric field may enhance the superconducting gap and critical current~\cite{eliashberg1970film,ivlev1971influence,tikhonov2018superconductivity}. 

In contrast, there has been less focus on nonlinear effects due to a combination of externally driven electric fields and currents. Notably, experiments on $s$-wave superconductor NbN observed a dc supercurrent to produce giant second-harmonic generation (SHG) in the THz range~\cite{nakamura2020nonreciprocal}. The supercurrent breaks inversion symmetry allowing for SHG. Separately, also in NbN, a dc super-current was seen to enhance the optical response at the edge of the superconducting gap, where peaks in the dissipative and reactive parts of the optical conductivity were observed~\cite{nakamura2019infrared}. 

In this work we analyse current enabled optical response in a superconductor. We study the effect of a uniform, dc supercurrent density $\Jsc$ on the ac optical conductivity $\sigma(\omega)$ in a time reversal symmetric superconductor at temperature $T=0$. This dc supercurrent may be due to an external source, or a screening current induced by a magnetic field~\cite{zhu2021discovery}. In clean systems, where the gap exceeds the scattering rate $2\Delta > \tau^{-1}$, we show the supercurrent to enable a large optical absorption peak of height $\propto j_\text{sc}^2$ at the gap edge $\omega = 2 \Delta$. This peak is due to single photon processes which excite quasiparticles across the gap. Naively, one might expect this peak to persist even for $\Jsc = 0$. However, when time reversal symmetry is present, this response is absent due to selection rules between quasiparticle modes at the gap edge, which cause the optical absorption to go continuously to zero at this point~\cite{bardeen1957theory,schrieffer2018theory,tinkham2004introduction,zimmermann1991optical,dressel2013electrodynamics,ahn2021theory}. In order to obtain a large gap edge optical response, time reversal symmetry must be broken. As superconductors with intrinsically broken time reversal symmetry are rare, we consider extrinsic breaking due to a finite supercurrent. 

A second important ingredient is current relaxation. The large gap edge response is thus absent in common minimal models of superconductors in which the current is conserved due to Galilean invariance. We provide a minimal model in the form of a multiband superconductor, as has been proposed for NbN~\cite{matsunaga2017polarization,tsuji2020higgs} and experimentally reported in $\text{NbSe}_2$~\cite{noat2010signatures,guillamon2008intrinsic,rodrigo2004stm,noat2015quasiparticle}, $\text{MgB}_2$~\cite{szabo2001evidence,bouquet2001specific,tsuda2001evidence,gonnelli2002direct,souma2003origin}, and $\text{FeSe}$~\cite{dong2009multigap,jiao2017superconducting}, or as may be realised via the proximity effect at a superconductor-metal interface.

\begin{figure}[t!]
    \centering
    \includegraphics[width=\linewidth]{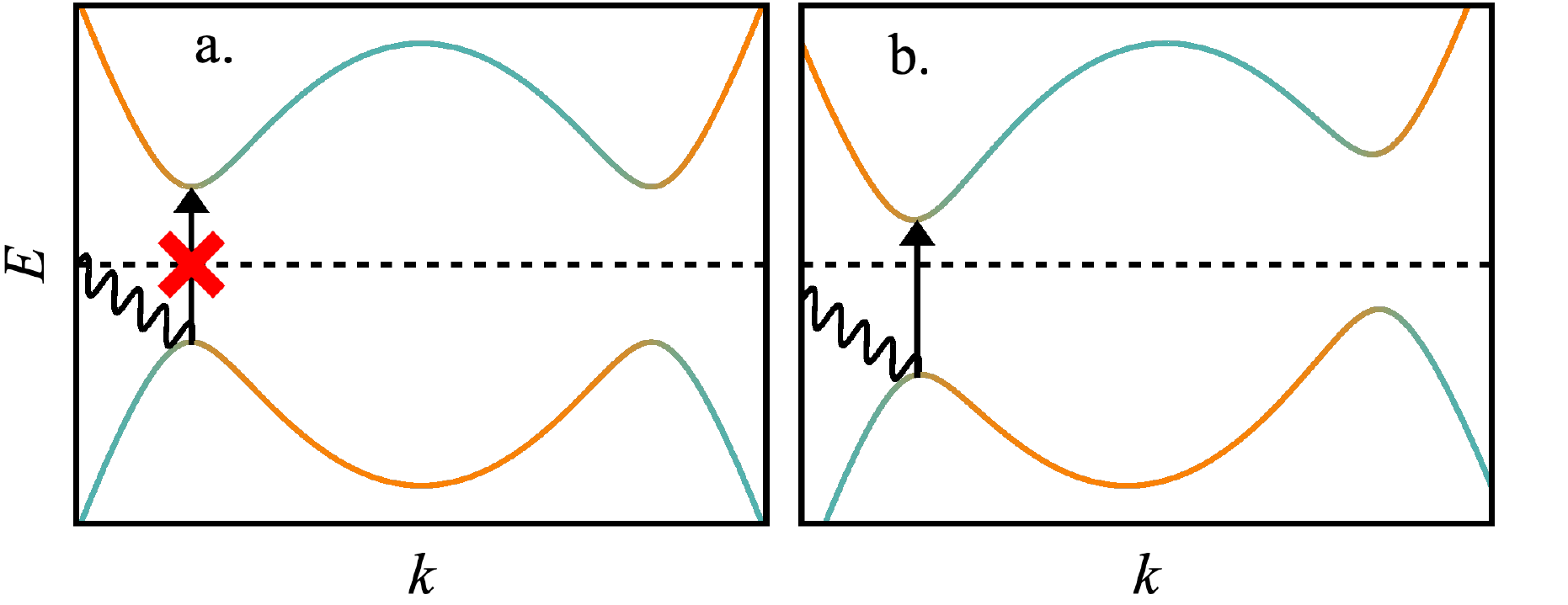}
    \caption{
    \emph{Optical absorption in time reversal symmetric superconductors}: The quasiparticle bandstructure with electron (hole) orbitals coloured orange (blue) a. time reversal symmetry leads to selection rules which prevents optical excitation of quasiparticles across the gap. b. A current breaks this symmetry, allowing excitation across the direct gap, resulting in an absorption peak at $\omega = 2\Delta$.
    }
    \label{Fig:bands}
\end{figure}

We focus on a response that is linear in the electric field, and thus characterised by the optical conductivity $\sigma(\omega)$
\begin{equation}
    \cexp{\J(\omega)} = \sigma(\omega) \E(\omega).
\end{equation}
We calculate the tensor $\sigma(\omega)$ in the presence of a dc supercurrent, for finite frequencies, via the Kubo formula
\begin{equation}
     i \omega \sigma_{ab}(\omega) =  - D_{ab} + i V \int_0^\infty dt \, \e^{i \omega t} \cexp{[ j_{a}(t), j_{b}(0)]}_0
     \label{eq:kubo_cond}
\end{equation}
where $\hJ$ is the current density operator, $V$ is the system volume, $a,b$ are spatial indices, $\cexp{\cdot}_0$ denotes thermal expectation values calculated for $\E(\omega) = 0$, and $D_{ab}$ is the Drude weight in the clean non-interacting limit (for a parabolic band $D_{ab} = n e^2 \delta_{ab}/m$ where $n$, $m$ and $e$ are electron density and mass, and charge). 

\emph{Current conserving system}---We begin with the canonical model of an $s$-wave superconductor: a single 
{\it parabolic} band with a contact interaction:
\begin{equation}
    \hH = \sum_{\k\sigma} \xi_\k c_{\k\sigma}^\dagger c_{\k\sigma} + g \sum_{\q} P_\q^\dagger P_\q
    \label{eq:H_galilean}
\end{equation}
where the interaction is attractive $g<0$, $\xi_\k = k^2/2m - \mu$ is the dispersion, and $P_\q^\dagger = \sum_{\k} c_{\k+\q \up}^\dagger c_{-\k\dn}^\dagger$ creates a pair of electrons with total momentum $\q$.

The optical response of the system~\eqref{eq:H_galilean} is trivial and independent of the presence of the supercurrent or otherwise. This follows from the Galilean invariance of $H$, which entails that the current density operator is a conserved integral of motion
\begin{equation}
    \hJ = \frac{e}{V} \sum_{\k\sigma} \v_\k c_{\k\sigma}^\dagger c_{\k\sigma} \quad \implies \quad [H,\hJ] =0
    \label{eq:J_exact}
\end{equation}
where $\v_\k =  \partial_\k \xi_{\k} = \k/m$ and $V$ is the system volume. Current conservation forbids the electromagnetic energy dissipation, and results in a trivial optical conductivity. Specifically, the real part of the optical conductivity is zero at all finite frequencies, while the imaginary part, given by $\sigma(\omega)=i D/ \omega$, encodes the uniform acceleration of the electron centre of mass by the electric field. 

Let us now examine how the trivial optical conductivity of a current conserving superconductor may be correctly obtained in mean-field theory when the system is carrying a supercurrent $\Jsc$. For $\Jsc \neq 0$, Cooper pairing occurs at a finite momentum $\q$, this is related to the supercurrent density $\Jsc$ and condensate velocity $\v_\mathrm{c}$ by
\begin{equation}
    \Jsc = e n \v_\mathrm{c}, \quad \,\, \v_\mathrm{c} = \q/2m.
\end{equation}
The corresponding mean field Hamiltonian, which depends on the Cooper pair momentum $\q$, is obtained in the usual way 
\begin{equation}
    \tH = \sum_{\k\sigma} \xi_{\alpha\k} c_{\k\sigma}^\dagger c_{\k\sigma} + 
   \Delta (\e^{i \theta} P_\q + \e^{-i \theta} P_\q^\dagger).   
    \label{eq:Hmf_1band}
\end{equation}
where $\Delta$ and $\theta$ are the magnitude and phase of the superconducting order parameter respectively. In the standard BCS theory, the phase factor $\e^{i \theta}$ is treated as a $c$-number. Although this leads to an artefactual violation of electron number conservation $[\tH,N] \neq 0$ ($N =\sum_{\k\sigma} c_{\k\sigma}^\dagger c_{\k\sigma}$), this treatment still allows for accurate calculation of the optical conductivity in clean and dirty systems at $\Jsc = 0$. The standard approach works for $\q=0$ as the pairing and kinetic terms in $\tH$ still conserve current as desired. However, for a finite supercurrent $\Jsc \neq 0$, Cooper pairing occurs at finite momentum $\q$, then the pair operators $P_\q$ in the mean field Hamiltonian alter the current $[P_\q,\J] \neq 0$. In this case treating the condensate phase $\e^{i\theta}$ as a $c$-number leads to an additional artefactual violation of the current conservation: $[\tH , \J] \neq 0$ and consequently artefactual contributions to the optical conductivity. 

In order to obtain the optical conductivity correctly, we treat $\theta$ as an operator associated with the phase of the condensate, the canonically conjugate operator $- i \partial_{\theta}$ then counts the number of Cooper pairs making up the condensate. The physical picture \cite{fu2010electron, vijay2016teleportation} is that $e^{i\theta}$ increases the number of Cooper pairs by one, while $P_\q$ decreases the number of  fermions by two. In this modified mean field approach, the total charge consists of both charge-$e$ fermions and charge-$2e$ Cooper pairs. Thus, the total charge number operator takes the form 
\begin{equation}
    \tN =  - 2 i \partial_{\theta} + \sum_{\k\sigma} c_{\k\sigma}^\dagger c_{\k\sigma} .
    \label{eq:Nmf_gal}
\end{equation}
Importantly this restores the conservation of electron number to the mean field theory, so that $[\tH,\tN]=0$, respecting the charge conservation present in the original theory~\eqref{eq:H_galilean}

This modified mean field approach also restores the current conservation of Galilean invariant superconductors. The total current consists of the currents carried by unpaired electrons and the condensate
\begin{equation}
\begin{aligned}
    \tJ & = \frac{e}{V} \Big(-  2 i\v_\mathrm{c}  \partial_{\theta} +   \sum_{\k\sigma} \v_\k  c_{\k\sigma}^\dagger c_{\k\sigma}\Big)
    \\
    & = \frac{e}{V} \Big( \v_\mathrm{c} \tN +   \sum_{\k\sigma} (\v_\k - \v_\mathrm{c})  c_{\k\sigma}^\dagger c_{\k\sigma} \Big)
    \label{eq:Jmf_gal}
\end{aligned}
\end{equation}
where in the second line, we have substituted~\eqref{eq:Nmf_gal} to eliminate $\partial_{\theta}$. From~\eqref{eq:Jmf_gal} one obtains $[\tJ,\tH] = 0$ as required.

We have presented physical arguments for the form of the number and current density operators appropriate for a superconducting mean field theory with a finite supercurrent. Importantly, these operators provide the correct book-keeping for the charge and current carried by the condensate. In the Galilean invariant system considered, this restores conservation of current to the mean field theory, and hence recovers the desired trivial optical conductivity. This approach is formalised in App.~\ref{app:Gauge_constraint} where we show the mean field current and number operators~\eqref{eq:Nmf_gal} and~\eqref{eq:Jmf_gal} may be obtained by extending the Hilbert space, and enforcing charge conservation as a gauge constraint which must be respected by the mean field theory. In App.~\ref{app:galilean_inv} we further show that in addition to conservation of charge, current and momentum this approach restores the correct Galilean transformation properties of $\tH$.

\emph{Lattice effects}---Current conservation in a Galilean invariant system follows from the fact that the interaction conserves current. However in solids, the presence of an underlying lattice means that current is generically not conserved: non-parabolic bands, Umklapp processes, or multi-band effects all break Galilean invariance, leading to current relaxation, and allowing for a non-trivial optical response. To illustrate this, we now show how coupling between bands with different effective masses allows for a non-trivial optical response in a current carrying superconductor. 

We consider a minimal model which, after the kinetic term has been diagonalised, consists of two parabolic bands $\alpha=1,2$, and interactions that scatter electrons within ($g_{11}$, $g_{22}$) and between ($g_{12}= g_{21}^*$) bands
\begin{equation}
    H = \sum_{\alpha\k\sigma} 
    \xi_{\alpha\k}c_{\alpha\k\sigma}^\dagger c_{\alpha\k\sigma} + \sum_{\k\alpha\beta} g_{\alpha\beta} P_{\alpha\q}^\dagger P_{\beta\q}
    \label{eq:H_2band}
\end{equation}
where $\xi_{\alpha\k} = k^2/2 m_\alpha - \mu_\alpha$, and $P_{\alpha\q}^\dagger = \sum_{\k} c_{\alpha\k+\q \up}^\dagger c_{\beta-\k\dn}^\dagger$. For simplicity, we assume the primary pairing interaction occurs in the first band, and induces a pairing potential in the second band via the inter-band coupling, providing the dominant source of pairing in both bands
\begin{equation}
    |g_{11}\cexp{P_{1\q}}| \gg |g_{12}\cexp{P_{2\q}}|, \quad |g_{12}\cexp{P_{1\q}}| \gg |g_{22}\cexp{P_{2\q}}|.
    \label{eq:Pairing_assumption}
\end{equation}
This may occur either due to a hierarchy of scales in the coupling $g_{11}\gg g_{12}\gg g_{22}$, or because the density of states at the Fermi surface is much greater in the first band. In this regime the mean field Hamiltonian takes the form
\begin{equation}
    \tH = \sum_{\alpha\k\sigma} 
    \xi_{\alpha\k} c_{\alpha\k\sigma}^\dagger c_{\alpha\k\sigma} + \sum_\alpha \Delta_\alpha\big( \e^{i \theta}  P_{\alpha\q}  + \mathrm{h.c.} \big)
    \label{eq:Hmf_int}
\end{equation}
where the primary gap is $\Delta_1 \e^{i\theta} = g_{11} \cexp{P_{1\q}^\dagger}$, and a smaller gap is induced in the second band $\Delta_2 = \Delta_1 g_{12}/g_{11}$. In Boguliobov--de Gennes (BdG) form we obtain
\begin{subequations}
\begin{equation}
    \tH = \sum_{\alpha\k} \psi_{\alpha\k}^\dagger \tH_{\alpha\k} \psi_{\alpha\k}
\end{equation}
where $\psi_{\alpha\k}^\dagger = (c_{\alpha\q/2+\k\up}^\dagger,c_{\alpha\q/2-\k\dn})$ is the Nambu spinor and we have defined the BdG matrices
\begin{equation}
\begin{aligned}
    \tH_{\alpha\k} & = \bxi_{\alpha\k} \tz + \delta \xi_{\alpha\k} \tn +   \Delta_\alpha \left( \e^{i\theta} \tp + \mathrm{h.c.}\right),
\end{aligned}
\end{equation}
\label{eq:h_bdg_int}
\end{subequations}
where $\bxi_{\alpha\k}  = \tfrac12 (\xi_{\alpha,\q/2 + \k}+\xi_{\alpha,\q/2 - \k}) $, $\delta \xi_{\alpha\k} =\tfrac12 (\xi_{\alpha,\q/2 + \k} - \xi_{\alpha,\q/2 - \k}) $, and $\tau_a$ are the usual Pauli matrices.

In this model Galilean invariance is broken, thus the current density is not conserved, but relaxes due to pair scattering between the bands. This non-conservation is quantitatively captured by the commutator of the current operator~\eqref{eq:J_exact} and Hamiltonian~\eqref{eq:H_2band} which details the rate of change of current due to each microscopic process
\begin{equation}
    [ \J , H ] = \delta \vec{j} \Big(g_{12} P_{11\q}^\dagger P_{22\q} - \mathrm{h.c.}\Big), \quad \delta \vec{j} = \frac{e \v_\mathrm{c} \delta m}{m_2 V},
    \label{eq:commJH_exact}
\end{equation}
where $\delta m = m_1 - m_2$, and $\v_\mathrm{c} = \q/2m_1$ is the condensate velocity. Though for the parabolic bands considered here current is conserved upon \emph{intra}-band scattering, the current changes upon \emph{inter}-band scattering: pairs of electron from the second band are scattered with rate $g_{12}$ into the first band, each causing a change to the current density of $\delta \J$, and \emph{vice versa}. The mean-field theory current operator must preserve this commutator in order to preserve the microscopic dynamics of current, correctly capture the current relaxation, and allow for accurate calculation of $\sigma(\omega)$.

In direct generalisation of ~\eqref{eq:Nmf_gal} and~\eqref{eq:Jmf_gal} we use the mean field current and number operators
\begin{equation}
\begin{aligned}
    \tJ & = \frac{e}{V} \Big( \v_{\mathrm{c}} \tN  +  \sum_{\alpha\k\sigma} (\v_{\alpha\k} - \v_{\mathrm{c}}) c_{\alpha\k\sigma}^\dagger c_{\alpha\k\sigma} \Big).
    \\
    \tN & =  - 2 i \partial_{\theta} + \sum_{\alpha\k\sigma} c_{\alpha\k\sigma}^\dagger c_{\alpha\k\sigma}  .
    \label{eq:Jmf_2band}
    \end{aligned}
\end{equation}
The mean field operators have the commutator
\begin{equation}
    [ \tJ , \tH ] = \frac{e \v_\mathrm{c} \delta m}{m_2 V} \Big(\Delta_2 \e^{i\theta} P_{22\q} - \mathrm{h.c.}\Big).
    \label{eq:Jmf_int}
\end{equation}
This commutation relation corresponds directly to the mean field equivalent of~\eqref{eq:commJH_exact}. Precisely, one obtains~\eqref{eq:Jmf_int} by making the mean field replacement to both sides of~\eqref{eq:commJH_exact}. Note that~\eqref{eq:commJH_exact} is symmetric upon exchanging the band labels, whereas~\eqref{eq:Jmf_int} is not, this is a consequence of our assumption that both the primary gap $\Delta_1$ and induced gap $\Delta_2$ arise due to pairing in the first band~\eqref{eq:Pairing_assumption}. Note also that, as required, the mean field current $\tJ$ is conserved in the limits of $m_1 = m_2$ and $g_{12} = 0$ in which Galilean invariance is restored. In BdG form we have
\begin{equation}
    \begin{gathered}
    \tJ  =e \v_\mathrm{c}\tN /V + \sum_{\alpha\k} \psi_{\alpha\k}^\dagger \tJ_{\alpha\k} \psi_{\alpha\k} ,
    \\
   \tJ_{1\k}  = e\bar{\vec{v}}_{1\k} \tn/V, \quad \tJ_{2\k} = e\left[ \bar{\vec{v}}_{2\k} \tn + \v_\mathrm{c} (\delta m/m_2) \tz \right]/V.
    \label{eq:J_2band}
    \end{gathered}
\end{equation}
where
$\bar{\vec{v}}_{\alpha\k} = \partial_\k \bxi_{\alpha\k} = \k/m_\alpha$. We note the appearance of the $\tz$ term in $\tJ_{2\k}$ with a current dependent coefficient. This term measures the density of electrons in the second band. This term will be important to obtaining a large gap edge response, and is only possible in a multi-band system where electrons may be exchanged between bands. Within a single band system the conservation of the total number of electrons prevents the coupling of electron density to an external field.

\begin{figure}[t!]
    \centering
    \includegraphics[width=0.95\linewidth]{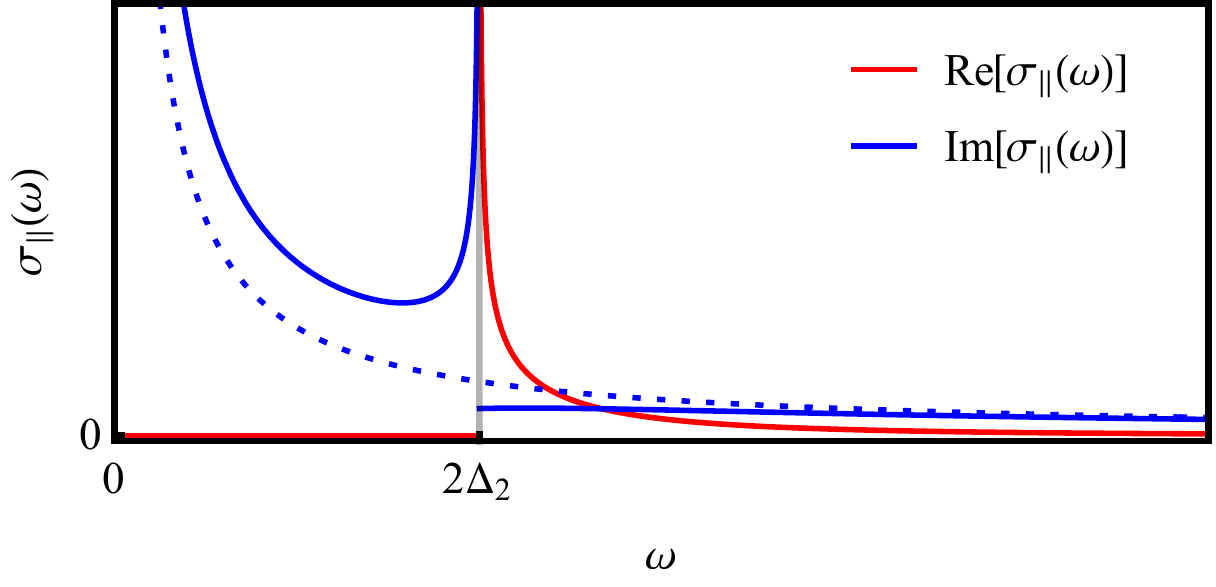}
    \caption{
    \emph{Current induced optical conductivity of a clean superconductor:} The red (blue) line show the dissipative (reactive) response~\eqref{eq:sigma_parr}. The dashed line shows the response for zero current $\q = 0$. The clean reactive response is shown (dashed) for comparison. Shown for a parabolic dispersion with parameters $(v_\mathrm{c} \delta m/ v_{2\mathrm{F}}m_2)^2 = 1/2$.
    }
    \label{Fig:clean}
\end{figure}

We may now evaluate the optical conductivity using~\eqref{eq:h_bdg_int} and~\eqref{eq:J_2band}. As both $\tN$, and $\e^{i \theta}$ commute with the mean field Hamiltonian $\tH$, we work in their diagonal basis, replacing both operators with their eigenvalues, allowing the current-current correlator to be evaluated using the usual fermionic algebra. In the clean system which we consider first, only the latter ($\propto \tz$) term of~\eqref{eq:J_2band} gives rise to a nontrivial response. To obtain this response, let us set the $z$-direction parallel to $\q$, due to axial symmetry the optical conductivity then has two independent components $\sigma_\parallel(\omega) = \sigma_{zz}(\omega)$, and $\sigma_\perp(\omega) = \sigma_{xx}(\omega) = \sigma_{yy}(\omega)$. In the presence of a supercurrent the perpendicular component remains trivial $\Re[\sigma_\perp(\omega)] = 0$, whilst the parallel component takes the form (derivation in App.~\ref{app:sigma_deriv})
\begin{equation}
    \Re [ \sigma_{\parallel}(\omega)] = \frac{\rho_{2\mathrm{F}}}{\omega}\!\! \left(\!\frac{e  v_{\mathrm{c}} \delta m}{m_2}\!\right)^{\!2}\!\!f\!\left(\!\frac{\omega}{2\Delta_2}\!\right)\!,\, \,\, f(x) = \frac{\pi \theta(x^2-1)}{x\sqrt{x^2-1}}
    \label{eq:sigma_parr}
\end{equation}
where we have set temperature $T=0$, $\theta(x)$ is the usual step function, and $\rho_\mathrm{2F}$ is the normal phase Fermi surface density of states of the second band. The appearance of mass in this result is due to the dependency on the curvature of the bands at the Fermi surface. 
As is necessary, this current mediated contribution to the optical conductivity~\eqref{eq:sigma_parr} disappears if the condensate velocity is zero $v_\mathrm{c} = 0$, or in either of the Galilean invariant cases of $m_1 = m_2$, or $g_{12} = 0$. The current mediated contribution is largest for in the case when the second band is much faster, $m_2 \ll m_1$, i.e. in the same regime as our assumption~\eqref{eq:Pairing_assumption}.

The current mediated optical response~\eqref{eq:sigma_parr} is most significant at the gap edge where it diverges. This divergence is mirrored in the reactive (imaginary) part of the optical conductivity (obtained via the Kramers-Kronig relation) which has an equivalent divergence as the gap is approached from below
\begin{equation}
    \Re [ \sigma_{\parallel}(2\Delta_2+\delta \omega)] \sim \Im [ \sigma_{\parallel}(2\Delta_2-\delta \omega)] = O( \delta\omega^{-1/2})
    \label{eq:div}
\end{equation}
where $\sim$ indicates asymptotic equality as $\delta\omega\to 0$. For fields oscillating at frequencies just below the gap, where the reactive part is large, a large ac current is induced, carried by electrons in the second band. As the peak is found only in $\sigma_\parallel$, with $\sigma_\perp$ remaining trivial, the induced ac current runs parallel to the super-current.

We emphasise that while the Doppler shift (the tilting of the dispersion) causes the indirect gap to shrink with increasing supercurrent $\omega_{\mathrm{ig}-} = 2 \Delta_2 - q k_{2\mathrm{F}}/2m_2 + O(q^2/m_2)$, the large response is always set by the current independent direct gap $ 2\Delta_2$. Moreover, as both the direct gap, and the change to the current~\eqref{eq:commJH_exact} induced by scattering $\delta \J$ are independent of position on the Fermi surface, the current induced response~\eqref{eq:sigma_parr} is obtained by integrating contributions from the entire Fermi surface. 

\begin{figure}[t!]
    \centering
    \includegraphics[width=\linewidth]{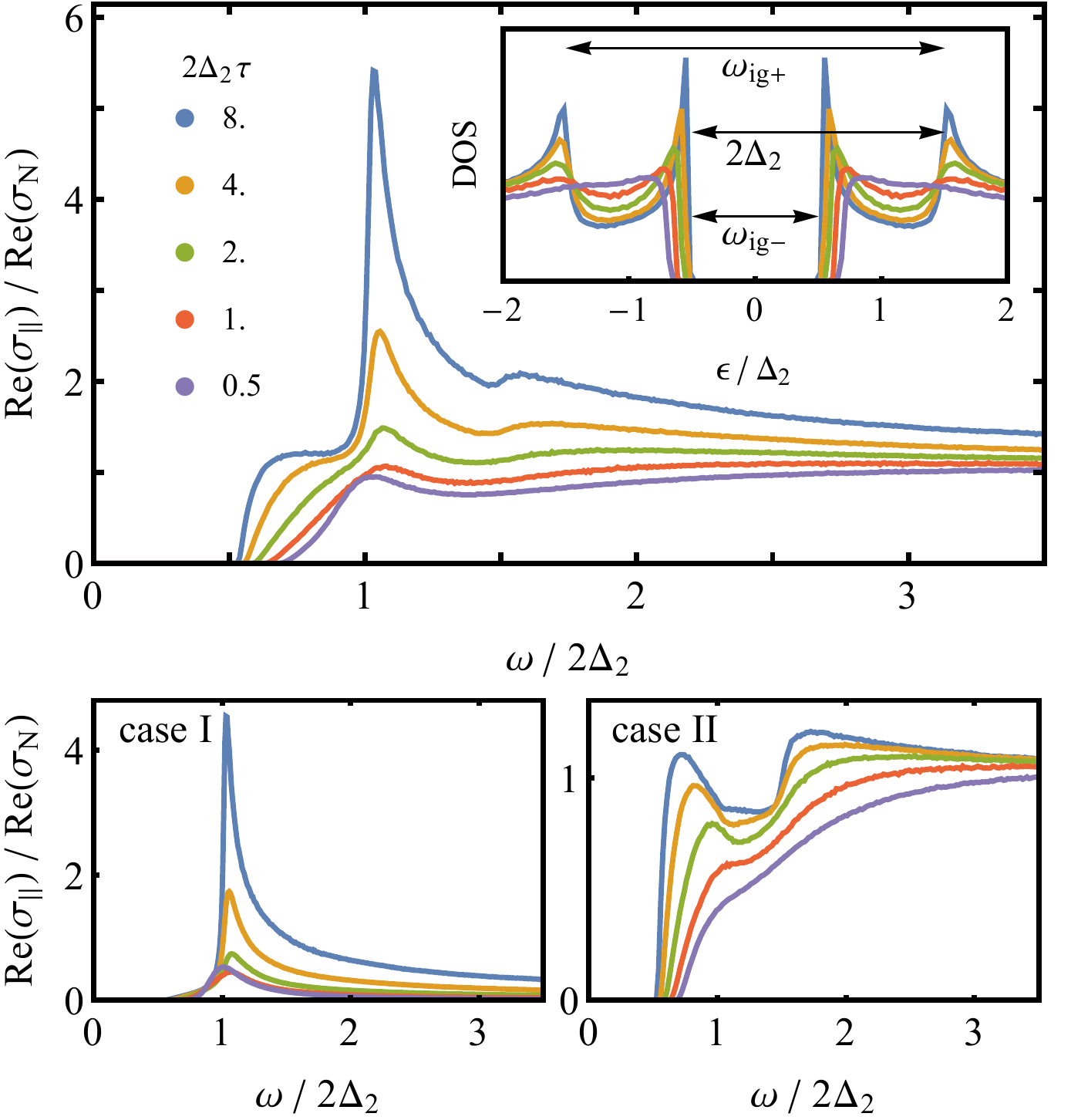}
    \caption{
    \emph{Current mediated response in a disordered superconducting wire:} The dissipative optical conductivity is shown in units of the normal phase dissipative response $\Re \sigma_\mathrm{N}$ for different scattering times (legend inset). The dissipative response onsets at the indirect gap $\omega_{\mathrm{ig}-}$, is peaked at the direct gap $2 \Delta$, and converges $\Re(\sigma_\mathrm{I}) / \Re(\sigma_\mathrm{N}) \sim 1$ far from the gap. The peak height becomes large $O(\tau^{3/2})$ in the clean limit ($2\Delta_2\tau \gtrsim 1$). The lineshape is a sum of case I and case II contributions are plotted separately in the lower panels. The gaps $\omega_{\mathrm{ig}-}$ and $2 \Delta_2$ are labelled on the quasiparticle particle density of states (inset, upper panel). Parameters: $(v_c \,\delta m/  m_2 v_\mathrm{2F})=0.3 $, $\delta\xi_{2\mathrm{F}} = q k_{2\mathrm{F}}/2m_2 = \Delta_2 /2$, $2 \Delta_2 \tau$ given in legend.
    }
    \label{Fig:disorder}
\end{figure}

\emph{Impurity scattering}---Impurities couple different momentum sectors, altering the optical response. We consider the effect of quenched non-magnetic disorder
\begin{equation}
    \tH \to \tH + \sum_{\k\q\sigma} V_{\k\q} c_{\k\sigma}^\dagger c_{\q\sigma},
    \label{eq:disorder}
\end{equation}
and focus on its effect on the dissipative response, from which the reactive response follows via the Kramers-Kronig relation.

Let us recall some useful properties from the usual BCS theory of response functions for the case $\Jsc = 0$, $T=0$. In the presence of time reversal symmetry, response functions can be decomposed into \emph{case I} (\emph{case II}) terms which are generated by operators which are even (odd) under time reversal symmetry~\cite{bardeen1957theory,tinkham2004introduction,schrieffer2018theory}. In the usual Nambu basis $\psi_{\alpha\k}^\dagger = (c_{\k\up}^\dagger,c_{-\k\dn})$ these correspond to BdG matrices $\tz$ ($\tn$) respectively. At zero temperature, case I and case II responses exhibit markedly different behaviour at the gap edge. To illustrate this consider an example case I (case II) operator $O^+$ ($O^-$) with BdG matrices $o_{\k\p}^{+} = \tau_0$, ($o_{\k\p}^{-} = \tau_3$) coupling the $\k$ and $\p$ momentum sectors, we obtain matrix elements
\begin{equation}
    |\bra{\epsilon_\k} o_{\k\p}^{\pm} \ket{\epsilon_\p}|^{2} = \frac{\epsilon_\k \epsilon_\p \mp \Delta_2^2 - \!\sqrt{\!(\epsilon_\k^2 - \Delta_2^2)(\epsilon_\p^2 - \Delta_2^2)\!}}{2\epsilon_\k \epsilon_\p}
\end{equation}
where $\ket{\epsilon_\k}$ is the eigenvector of $\tH_{2,\k}$ for zero current $\delta \xi_{2,\k} = 0$. In the limit of states either side of the gap edge $(\epsilon_\k, \epsilon_\p ) \to (\Delta_2, -\Delta_2)$ we find that in case II $|\bra{\epsilon_\k} o_{\k\p}^{-} \ket{\epsilon_\p}|^2 \to 0$, whereas in case I the matrix element $|\bra{\epsilon_\k} o_{\k\p}^{+} \ket{\epsilon_\p}|^2$ approaches double its normal phase ($\Delta_2 = 0$) value. Thus we see that, due to selection rules (or equivalently coherence factors), the case I (case II) response is enhanced (suppressed to zero) as $\omega \to 2 \Delta_2^+$. Whereas inside the gap $|\omega| < 2 \Delta_2$ both the case I and II responses are zero. Thus the (enhanced) case I response drops discontinuously to zero at $\omega = 2\Delta_2$. The distinct gap edge behaviour of case I and case II operators relies only on the time reversal symmetry of $H$~\cite{anderson1959theory}, and thus holds in the presence of non-magnetic disorder. Despite the possibility of enhanced linear response at the gap edge, no such large gap edge response is exhibited in the optical conductivity, as the current operator is odd under time reversal symmetry, and thus purely case II~\cite{bardeen1957theory,schrieffer2018theory,tinkham2004introduction,mattis1958theory,abrikosov1959theory,nam1967theory,zimmermann1991optical,dressel2013electrodynamics,chen1992transport,chen1993theory,lee2018optical}.

In the presence of the current this picture is altered, the pairing is at finite momentum, and thus the BdG matrices $\tz$ and $\tn$ are not odd/even under time reversal. However, for weak scattering, the distinction in behaviours of the matrix elements close to the gap remains, and so we retain the nomenclature of case I and case II for $\tz$ and $\tn$ respectively. Using this convention, we find the case I term in the current~\eqref{eq:J_2band} leads directly to a case I term in the optical conductivity~\eqref{eq:kubo_cond}, in addition to the usual case II term
\begin{equation}
    \sigma(\omega) = \sigma_\mathrm{I}(\omega) + \sigma_\mathrm{II}(\omega),
    \label{eq:sigma_I_II}
\end{equation}
In~\eqref{eq:sigma_I_II} the case I/II cross terms are omitted as they disorder average to zero (see App.~\ref{eq:Cross_term}).
Moreover, the optical conductivity is further altered by the current induced Doppler shift, which causes the indirect (momentum non-conserving) gap $\omega_{\mathrm{ig\pm}} = 2 \Delta_2 \pm q k_{2\mathrm{F}}/2m_2 + O(q^2/m_2)$ to become smaller than the direct (momentum conserving) gap $2 \Delta_2$ (see Fig~\ref{Fig:bands} and Fig~\ref{Fig:disorder} inset).

We now discuss the case I and case II contributions to the optical conductivity in the presence of a finite current. These are obtained via exact diagonalisation in Fig~\ref{Fig:disorder}, where the scattering term~\eqref{Fig:disorder} is used for Gaussian distributed elements $V_{\k,\q} = V_{-\k,-\q}^*$ (numerical details in App.~\ref{app:ED}). The current mediated case I contribution consists of a single peak at $2 \Delta$. Unlike in the clean case, where the response is zero at frequencies below the peak
$\omega<2 \Delta_2$, in the presence of scattering a tail extends down to the indirect gap at $\omega_{\mathrm{ig-}} < 2 \Delta_2$. Most significantly, the presence of a current means the sharp peak at the direct gap $\omega = 2 \Delta_2$ is not protected by time-reversal symmetry, and becomes rounded. This effect is quantitatively captured by repeating the calculation of~\eqref{eq:sigma_parr} with a finite quasiparticle lifetime $\tau$ (see App.~\ref{app:sigma_deriv}) yielding
\begin{equation}
\begin{aligned}
    \sigma_{\mathrm{I},\parallel}(\omega) = \frac{\rho_{2\mathrm{F}}}{\omega}\!\! \left(\!\frac{e  v_{\mathrm{c}} \delta m}{m_2}\!\right)^2 \!f\!\left(\frac{\omega \tau - i }{2 \Delta_2 \tau}\right)\!, \,\,  f(z) = \!\frac{2 i \arcsin(z)}{z \sqrt{1 - z^2 }}.
    \end{aligned}
    \label{eq:finite_tau_cond}
\end{equation}
We find~\eqref{eq:finite_tau_cond} the clean divergences in the dissipative and reactive parts both become truncated at a maximal height $\sigma(\omega) = O(\tau^{1/2})$. This may be compared with the normal phase response at the same frequency $\Re (\sigma_\mathrm{N}) = D \tau / (1 + \omega^2 \tau^2)$~\cite{lee1985disordered}. In the clean limit $2 \Delta_2 \tau \gtrsim 1$, at the gap edge $\omega = 2\Delta_2$, we thus find a peak contrast of
\begin{equation}
    \frac{\Re(\sigma_{\parallel})}{ \Re(\sigma_\mathrm{N})} \sim \frac{\pi \rho_{2\mathrm{F}}}{2 D}\!\! \left(\!\frac{e  v_{\mathrm{c}} \delta m}{m_2}\!\right)^{\!\!2} \!(2 \Delta \tau)^{\tfrac32}
    = \frac{\pi d}{2}\!\! \left(\! \frac{v_c \delta m}{v_{2\mathrm{F}} m_2} \!\right)^{\!\!2} \!(2 \Delta \tau)^{\tfrac32}
    \label{eq:scaling}
\end{equation}
where $\sim$ indicates asymptotic equality at large $\tau$, and the second equality in~\eqref{eq:scaling} applies for a parabolic band in dimension $d$ with Fermi velocity $v_{2\mathrm{F}}$. We find the $O(\tau^{3/2})$ scaling is in good agreement with numerics (see Fig.~\ref{Fig:disorder} and App.~\ref{app:ED}). Lastly we note that, as before~\eqref{eq:div}, the real and imaginary peaks are asymptotically equal in this limit $\Re(\sigma_{\parallel}) \sim \Im(\sigma_{\parallel})$.

The case II contribution (Fig~\ref{Fig:disorder}) exhibits no dramatic changes from the $\Jsc = 0$ case, altered only due to the change in the gap structure: $\sigma_\mathrm{II}$ converges to the normal phase Drude response far above the gap $\Re(\sigma_\mathrm{II}) / \Re(\sigma_\mathrm{N}) \sim 1$, and remains of comparable scale at all frequencies above the indirect gap. In the absence of a current, the case II response goes continuously to zero at the gap edge $\omega = 2 \Delta_2$. In the presence of the supercurrent this divides into three onsets, one at each of the gaps---in one spatial dimension the indirect gaps are linear onsets, whereas the direct gap is softer due to additional suppression of the matrix elements. In larger spatial dimension the onsets at the indirect gaps become softer, as only parts of the Fermi surface where $\k$ is parallel to $\q$ contribute to the density of states at the indirect gap. The case I and case II contributions are shown separately in the lower panels of Fig~\ref{Fig:disorder}.


\emph{Discussion}---In this manuscript we have described a supercurrent enabled optical response present in BCS superconductors. Such a supercurrent may be due to an external current source, or a screening current induced by a magnetic field. An important ingredient was the breaking of Galilean symmetry, i.e. the presence of some form of lattice physics. As an example we considered pair scattering between bands with different Fermi velocities. However, the type of lattice physics employed is unimportant, the current mediated response should be expected in any system where electron scattering does not conserve the total current. Other examples of physics which provide such non-conservation of current include Umklapp scattering, and strongly non-parabolic dispersion at the Fermi surface. These examples, though generically providing a weaker response, may be appealing avenues for future study as they require only a single band.

The current enabled optical response predicted in this work is in good agreement with the observations of Ref.~\cite{nakamura2019infrared}, where a current enabled change to the optical conductivity was reported, which manifested in an absorption (reactance) peak beginning at $\omega = 2\Delta$ and extending just above (below) the gap. These peaks were observed to reach a height quadratic in the current, in agreement with~\eqref{eq:finite_tau_cond}. Our theory differs from the alternative theoretical proposal of Ref.~\cite{moor2017amplitude}, which reported a current enabled response in a dirty single band BCS superconductor. In contrast we find that disorder alone does not lead to a large response at the gap edge. 

We comment on interesting avenues for future studies. The large reactive response which is generated in the superconducting gap provides a mechanism for the coherent coupling of THz radiation and superconducting circuits~\cite{lauk2020perspectives}. Additionally, we note that the current enabled optical response allows for excitation of the Higgs mode. The coherent generation of excitations above the superconducting gap results in a suppression of the pairing potential, this perturbation decays amidst long lived oscillations of the collective `Higgs' mode $\delta\Delta(t) \sim \cos(2 \Delta t)/\sqrt{2 \Delta t}$~\cite{volkov1973collisionless,matsunaga2013higgs,murotani2019nonlinear,bellitti2021incoherent,shimano2020higgs}. Here we present a single photon process, second order in the current, which permits such a coupling. This provides an alternative mechanism to previous efforts which have focused on achieving such a coupling using multi-photon processes~\cite{shimano2020higgs}. Lastly we note that our theory relies only on the scattering between bands, which may be taken as parabolic in the simplest example. We leave to future investigations the exploration of how effects due to the quantum geometry of electrons~\cite{ahn2021theory,ahn2021riemannian,takasan2021current} may also contribute to the superconducting optical response.

\emph{Acknowledgements}---We are especially grateful for insightful discussions with Boris Spivak, and also to Chris Baldwin, Matteo Bellitti, Margarita Davydova, Chaitanya Murthy and Nisarga Paul. This work is supported by the U.S. Army Research Laboratory and the U.S. Army Research Office through the Institute for Soldier Nanotechnologies, under Collaborative Agreement Number W911NF-18-2-0048. LF was partly supported by the David and Lucile Packard Foundation.

\bibliographystyle{apsrev4-1}
\bibliography{SC_bib}

\appendix

\section{Gauge constraint derivation}
\label{app:Gauge_constraint}

In this appendix we formalise the arguments presented in the main text, showing how the current and number operators obtained in the main text~\eqref{eq:Jmf_gal} may be obtained from a BCS theory in an extended Hilbert space in which charge conservation is enforced using a gauge constraint. 

We begin with a many fermion Fock space $\cF$ with associated creation (annihilation) operators $f_{\alpha\k\sigma}^
\dagger$ ($f_{\alpha\k\sigma}$). We then extend this Hilbert space to include a degree of freedom which counts total number of particles in the system
\begin{equation}
    \cH_\mathrm{E} = \cF \otimes \cH_\phi
\end{equation}
where $\cH_\phi = L_2(S_1)$ is spanned by $\ket{n}, n \in \mathbb{Z}$, or equivalently in its dual basis $\ket{\phi}, \phi \in [0,2\pi]$ in the canonically conjugate basis. Operators acting on $\cH_\phi$ include the number operator $N_\phi = - i \partial_\phi$ and the raising (lowering) operator $\e^{i \phi}$ 
\begin{equation}
N_\phi \ket{n} = n \ket{n}, \qquad \e^{i \phi} \ket{n} = \ket{n+1}.
\end{equation}
We impose the gauge constraint that $N_\phi$ is equal to the total number of fermions $N = \sum_{\k\sigma} f_{\alpha\k\sigma}^\dagger f_{\alpha\k\sigma}$
\begin{equation}
	\cF ' = \mathrm{span}\left( \ket{\psi} : \ket{\psi}  \in \cH_\mathrm{E}, (N - N_\phi) \ket{\psi} = 0 \right).
\end{equation}
or in more concise terms, we fix to a gauge
\begin{equation}
	N = N_\phi
\label{eq:g1}
\end{equation}
One may verify that $\cF '$ is itself a fermionic fock space with associated fermionic creation (annihilation) operators
\begin{equation}
c_{\alpha\k\sigma}^\dagger = f_{\alpha\k\sigma}^\dagger \e^{i \phi}, \qquad c_{\alpha\k\sigma} = f_{\alpha\k\sigma} \e^{-i \phi}.
\label{eq:cf}
\end{equation}
more specifically, $\cF$ and $\cF'$ are isomorphic, related by the unitary map
\begin{equation}
	c_{\alpha\k\sigma}^\dagger = U f_{\alpha\k\sigma}^\dagger U^\dagger, \qquad U = e^{i \phi N}
\end{equation}

A useful picture for understanding this contruction is to view~\eqref{eq:cf} a decomposition of the electron $c_{\alpha\k\sigma}^\dagger$ into a chargeless fermionic quasiparticle $f_{\alpha\k\sigma}^\dagger$ and a charge $\e^{i \phi}$. In accordance with this intution, the electrons have the global $\mathrm{U}(1)$ transformation
\begin{equation}
	\e^{i \alpha Q} c_{\alpha\k\sigma}^\dagger \e^{-i \alpha Q} = c_{\alpha\k\sigma}^\dagger \e^{ i e \alpha}, \qquad Q = e N_\phi
\end{equation}
Due to our extension of the Hilbert space, there is a gauge freedom in defining operator on $\cF$
\begin{equation}
	O \to O' = O + \lambda ( N - N_\phi).
	\label{eq:equiv_ops}
\end{equation}
In the full theory this Gauge choice is inconsequential. However in the mean field theory, this gauge choice leads to ambiguities which must be resolved.

We now turn to consider the full Hamiltonian
\begin{equation}
    \hH = \sum_{\alpha\k\sigma} \xi_{\alpha\k} c_{\alpha\k\sigma}^\dagger c_{\alpha\k\sigma} + \sum_{\alpha\q} g_{\alpha\beta} P_{\alpha\q}^\dagger P_{\beta\q}
\end{equation}
where $P_{\alpha\q}^\dagger = \sum_{\k} c_{\alpha\k+\q \up}^\dagger c_{\alpha-\k\dn}^\dagger$ and $\xi_{\alpha\k} = k^2/2m_\alpha - \mu_\alpha$ creates a pair of electrons with total momentum $\q$. The corresponding number, momentum, and current density operators are given by
\begin{subequations}
\begin{align}
	N &= \sum_{\alpha\k\sigma} c_{\alpha\k\sigma}^\dagger c_{\alpha\k\sigma} 
\\
	K &= \sum_{\alpha\k\sigma} \k c_{\alpha\k\sigma}^\dagger c_{\alpha\k\sigma}
\\
	\J &= \frac{e}{V} \sum_{\alpha\k\sigma} \v_{\alpha\k} c_{\alpha\k\sigma}^\dagger c_{\alpha\k\sigma}
\end{align}
\end{subequations}
where $\v_{\alpha\k} = \partial_\k \xi_{\alpha\k}$. In a situation where electrons cannot flow in an out of the system from an external environment the electronic charge of the system remains fixed to its initial value $N_0$, thus we have the additional constraint
\begin{equation}
	N_\phi = N_0 .
\label{eq:g2}
\end{equation}

The BCS mean field analysis break conservation of number of fermions. Here we seek to apply this analysis in a manner that preserves the conservation of charge, specifically we relax the constraint~\eqref{eq:g1} (the conservation of chargless fermionic quasiparticles) whilst maintaining the constraint~\eqref{eq:g2} (the conservation of charge). This is achieved by applying the usual mean field arguments in the basis of the $f_{\alpha\k\sigma}^\dagger$, following the example used in the main text, we consider pairing dominated by the $\alpha$ band 
\begin{equation}
\begin{aligned}
    \cexp{ F_{1\k} } & = \delta_{\k\q}\Delta_2 / g_{2 1} = \delta_{\k\q} \Delta_1 / g_{1 1}
    \\
    \cexp{ F_{2\k} } & = 0
\end{aligned}
\end{equation}
where
\begin{equation}
    F_{\alpha\q}^\dagger = \sum_{\k} f_{\alpha\k+\q \up}^\dagger f_{\alpha-\k\dn}^\dagger.
\end{equation}
This leads to the usual mean field replacement in the contact interaction
\begin{multline}
    H_\mathrm{int} = \sum_{\alpha\q} g_{\alpha\beta} F_{\alpha\q}^\dagger F_{\beta\q}
    \\
    \to \tH_\mathrm{int} = \sum_{\alpha} ( \Delta_\alpha F_{\alpha\q} + \mathrm{h.c.}) - \Delta_1^2/g_{11}
\end{multline}
Neglecting the overall constant this yields the mean field Hamiltonian
\begin{equation}
\begin{aligned}
	\tH & = \sum_{\alpha\k\sigma} \xi_{\alpha\k} f_{\alpha\k\sigma}^\dagger f_{\alpha\k\sigma} +  \sum_{\alpha} \left(  \Delta_\alpha F_{\alpha\q} + \mathrm{h.c.} \right)
\end{aligned}
\end{equation}
which we may subsequently write in terms of the electron algebra
\begin{equation}
    \tH = \sum_{\k\sigma} \xi_{\alpha\k} c_{\alpha\k\sigma}^\dagger c_{\alpha\k\sigma} + 
   \sum_{\alpha} \Delta_\alpha (\e^{2i \phi} P_{\alpha\q} + \mathrm{h.c.})
\end{equation}
In relaxing the constraint on the number of fermionic quasiparticles, we introduce a physical distinction between previously gauge equivalent operators~\eqref{eq:equiv_ops}. It is thus necessary to fix a gauge consistent with the mean field theory. In cases where there are conservation laws, this gauge fixing is easily achieved by requiring that the mean field theory also respect the appropriate conservation laws. For example, in the case of momentum, we have $[H , \K ] = 0$, and we seek to find $\tK = \K + \vec{\lambda} ( N_\phi - N)$ such that $[\tH, \tK ] = 0$. One finds the solution
\begin{equation}
\begin{aligned}
	\tK & = \K + (\q /2 )( N_\phi - N) 
	\\
	& =  \q N_\phi /2 + \sum_{\k\sigma} (\k - \q/2) c_{\alpha\k\sigma}^\dagger c_{\alpha\k\sigma}
\end{aligned}
\end{equation}
Applying the same logic to the case of charge we obtain
\begin{equation}
	\tN = N_\phi.
\end{equation}
Current is generically not a conserved quantity, but we can nevertheless gauge fix in an analogous manner, requiring that the algebra of $[H,\J]$ is conserved. Specifically we want that as
\begin{equation}
    [\tJ,P_{\alpha\q}^\dagger] = e\v_{\alpha\q} P_{\alpha\q}^\dagger / V
\end{equation}
that its mean field replacement $P_{\alpha\q}^\dagger \to \cexp{P_{\alpha\q}^\dagger} = \delta_{\alpha1}\Delta_1 \e^{2i \phi}/g_{11}$ should act on the total current in an analogous way
\begin{equation}
    [\tJ,\e^{2i \phi}] = e\v_{\alpha\q} \e^{2i \phi} / V
\end{equation}
yielding the mean field current operator
\begin{equation}
    \tJ = \frac{e}{V} \Big( \v_{1\q} N_\phi + \sum_{\k\sigma} (\v_\k - \v_{1\q}) c_{\alpha\k\sigma}^\dagger c_{\alpha\k\sigma} \Big).
\end{equation}
It is natural in the superconducting context to identify the ``missing fermions'' with the condensate. Specifically, the number of Cooper pairs in the condensate is taken to be
\begin{equation}
    N_\theta = (N_\phi - N)/2
\end{equation}
with a canonically conjugate coordinate
\begin{equation}
    \e^{i \theta} = \e^{2i \phi}.
\end{equation}
With these substitutions the results in the main text~\eqref{eq:Hmf_1band},~\eqref{eq:Nmf_gal},~\eqref{eq:Jmf_gal} and~\eqref{eq:Jmf_2band} are readily obtained.

\section{Galilean invariance, conservation of momentum and charge conservation in the mean field theory}
\label{app:galilean_inv}

In this appendix we show the argument presented in the main text and App.~\ref{app:Gauge_constraint} restores the correct Galilean transformation properties of the mean field Hamiltonian $\tH$. Corollaries of the restored behaviour under Galilean transformations are the restored conservation of momentum, current and particle number.

First we recap the technical content of Galilean invariance. A Galilean transformation to a moving reference frame $\r \to \r' = \r + \v t$ is implemented by the unitary
\begin{equation}
    U_{\u} = \exp ( i \u \cdot \vec{g}), \quad \vec{g} = M \hR - \hK t
\end{equation}
where $\hK$ is the total momentum and $\hR$ the centre of mass
\begin{equation}
    \hK = \sum_{\k\sigma} \k c_{\k\sigma}^\dagger c_{\k\sigma}, \quad \hR = M^{-1}\sum_{\r\sigma} m\r c_{\r\sigma}^\dagger c_{\r\sigma}.
\end{equation}
and $M = m N$ is the total mass, with $N$ the number of electrons and $m$ their effective mass. Under this transformation position and momentum base creation operators transform as
\begin{equation}
\begin{aligned}
    U_\u c_{\k,\sigma}^\dagger U_\u^\dagger & =  c_{\k + m \u ,\sigma}^\dagger \exp [- i t \k \cdot \u - i m u^2 t/2 ]
    \\
    U_\u c_{\r,\sigma}^\dagger U_\u^\dagger & =  c_{\r + \v t,\sigma}^\dagger \exp [ i m \r \cdot \u + i m u^2 t/2]
\end{aligned}
\end{equation}
where

Galilean symmetry is the statement that the dynamics of $\ket{\psi}$ and $\ket{\psi'} = U_\u\ket{\psi}$ are generated by the same Hamiltonian
\begin{equation}
    i \partial_t \ket{\psi} = H \ket{\psi}, \quad i \partial_t (U_\u \ket{\psi}) = H U_\u \ket{\psi}.
\end{equation}
A system is thus Galilean symmetric if and only if the Hamiltonian satisfies
\begin{equation}
    \hH = U_{\u} \hH U_{\u}^\dagger - i U_{\u} \partial_t U_{\u}^\dagger.
    \label{eq:gal_trans}
\end{equation}
It is readily verified that the full Hamiltonian~\eqref{eq:H_galilean} satisfies the transformation rule~\eqref{eq:gal_trans} provided the dispersion is parabolic.

However we cannot make the requirement of Galilean invariance on the mean field Hamiltonian $\tH$. This is because the mean field theory explicitly breaks Galilean symmetry by privileging a specific frame, the rest frame of the condensate. Instead the Galilean transformation relates a family of mean field Hamiltonians $\tH_\q$ parameterised by the cooper pair momentum $\q$
\begin{equation}
    \tH_\q = \sum_{\k\sigma} \xi_\k c_{\k\sigma}^\dagger c_{\k\sigma} + 
   \Delta (\e^{i \theta_\q} P_\q + \mathrm{h.c}).   
    \label{eq:Hmf_1band_app}
\end{equation}
Here, for clarity, we have explicitly labelled the momentum $\q$ carried by the mean field, writing $\e^{i \theta_\q}$. As the cooper pair momentum transforms as $\q \to \q' = \q + 2 m \u $ under Galilean transformations, the mean field Hamiltonian must satisfy the transformation rule
\begin{equation}
    \tH_{\q + 2 m \u} = \tilde{U}_{\u} \tH_\q \tilde{U}_{\u}^\dagger - i \tilde{U}_{\u} \partial_t \tilde{U}_{\u}^\dagger.
    \label{eq:Hmf_Gal_trans}
\end{equation}

We now show that the mean field Hamiltonian $\hH$ satisfies the Galilean transformation rule~\eqref{eq:Hmf_Gal_trans} if we make the appropriate requirements for the transformation properties of the superconducting phase $\e^{i \theta_\q}$. In detail, if we define a mean Galilean transformation operator for the mean field theory $\tilde{U}_{\u} = \exp ( i \u \cdot \vec{\tilde{g}})$, where $\vec{\tilde{g}} = M \tR - \tK t$ where we require
\begin{equation}
    \begin{aligned}
        {[\e^{i \theta_\q},\e^{-i \theta_\q}]} & = 0, \quad & 
        {[\tN,\e^{i \theta_\q}]} & = 2\e^{i \theta_\q}, 
        \\
        {[\tR,\e^{i \theta_\q}]} & = - i \partial_\q \e^{i \theta_\q}, \quad &
    [\tK,\e^{i \theta_\q}] & = \q \e^{i \theta_\q}.
    \end{aligned}
\end{equation}
then we recover the desired property, that $\e^{i \theta_\q}$ transforms as a creation operator for a pair of electrons with total momentum $\q$
\begin{equation}
    U_\u \e^{i \theta_\q} U_\u^\dagger = \e^{i \theta_{\q + 2 m \u}} \exp [- i t \q \cdot \u - i m u^2 t ]
\end{equation}
and it is readily verified that the mean field Hamiltonian~\eqref{eq:Hmf_1band_app} satisfies the transformation rule~\eqref{eq:Hmf_Gal_trans}. 

It may be further verified that the mean field Hamiltonian conserves momentum and particle number
\begin{equation}
    [\tH_\q,\tK] = 0, \quad [\tH_\q,\tN] = 0, \quad [\tH_\q,\tJ] = 0
\end{equation}
where the current is given by $\tJ = - e \tK/m$.

\section{Derivation of the optical response in the effective one-band model}
\label{app:sigma_deriv}

In this appendix we derive the optical conductivity~\eqref{eq:sigma_parr} as plotted in Fig~\ref{Fig:bands}, and obtain the form~\eqref{eq:finite_tau_cond} for the case I response in the presence of a finite scattering time.

Starting the effective Hamiltonian~\eqref{eq:h_bdg_int} and current operator~\eqref{eq:J_2band}, given to first order in $(t/\bxi_{2\k})$ and zeroth order in $(\delta \xi_{1\k} - \delta \xi_{2\k})/ \bxi_{2\k}$ by
\begin{equation}
\begin{aligned}
    \tH_{2\k} & = \bxi_{2\k} \tz + \delta \xi_{2\k} \tn +   \Delta_2 \left( \e^{i\theta} \tp + \mathrm{h.c.}\right),
    \\
    \tJ_{2\k} & = e\left[ \bar{\vec{v}}_{2\k} \tn + \v_\mathrm{c} (\delta m/m_2) \tz \right]/V.
\end{aligned}
\end{equation}
The current correlator is then easily evaluated
\begin{equation}
    \cexp{[ J_{a}(t), J_{a}(0)]} = \sum_{\k \alpha\beta} (f_{\k\alpha} - f_{\k\beta})|\bra{\k\alpha}\tilde{j}_{a,2\k}\ket{\k\beta}|^2 \e^{i ( E_{\k\beta} - E_{\k\alpha})t}
\end{equation}
where $\alpha = 1,2$ enumerates the eigenvectors of the effective Hamiltonian $\tH_{2\k} \ket{\k\alpha} = E_{\k\alpha} \ket{\k\alpha}$, and $f_{\k\alpha} = f(E_{\k\alpha})$ is population of the $\alpha$th level, given by the Fermi-Dirac distribution. Evaluating this at $T=0$, including a finite quasiparticle lifetime $\tau>0$ and we obtain 
\begin{equation}
\begin{aligned}
    C(\omega) & = V \int_0^\infty dt \e^{i \omega t}\cexp{[ j_{a}(t), j_{a}(0)]} 
    \\
    & = \!\int_{-\infty}^\infty d \xi  \frac{ \rho_{2\mathrm{F}} j_{a\q}^2 \Delta_2^2 }{\Delta_2^2 + \xi^2} \sum_{s = \pm }  \frac{i s}{\omega + i /\tau - 2 s \sqrt{\xi^2 + \Delta_2^2}}
    \\
    & =  \int_{2\Delta}^\infty \frac{ 4 \rho_{2\mathrm{F}} j_{a\q}^2 \Delta_2^2 d \omega'}{\omega'\sqrt{{\omega'}^2 - 4 \Delta_2^2}}  \sum_{s = \pm }  \frac{i s}{\omega - s \omega' + i /\tau}
    \end{aligned}
    \label{eq:Cw}
\end{equation}
where $\rho_{1\mathrm{F}}$ is the normal phase density of states of the first band at the Fermi surface and we have set
\begin{equation}
    j_{a\q} = \frac{e q_a \delta m}{2 m_1m_2} = \frac{e v_{\mathrm{c},a} \delta m}{ m_2}
\end{equation}
where $\v_\mathrm{c} = \q/2m_1$ is the condensate velocity. To obtain $C(\omega)$ in the $\tau \to \infty$ limit we apply the Sokhotski–Plemelj theorem and evaluate the subsequent integrals to obtain
\begin{subequations}
\begin{equation}
    \sigma_{aa}(\omega) = \frac{i D_{aa}}{\omega} + \frac{e^2 \rho_{2\mathrm{F}}}{\omega} \cdot \left(\frac{ v_{\mathrm{c},a} \delta m }{ m_2} \right)^2 \cdot f\left(\frac{\omega}{2 \Delta_2}\right)
\end{equation}
where
\begin{equation}
    f(x) = -f^*(-x) = \begin{cases}
    \displaystyle \frac{\pi - 2 i \operatorname{arccosh}(x)}{x\sqrt{x^2 - 1}} \qquad & x > 1
    \\[10pt]
    \displaystyle \frac{2 i \arcsin(x)}{x \sqrt{1 - x^2 }} \qquad & 0 < x < 1
    \end{cases}
\end{equation}
\label{seq:sigma_clean}
\end{subequations}
Taking the real part we obtain~\eqref{eq:sigma_parr} as desired, both the real and imaginary parts are shown in Fig~\ref{Fig:bands}.

For finite $\tau$ we may evaluate~\eqref{eq:Cw} directly, obtaining
\begin{equation}
    \sigma_{aa}(\omega) = \frac{i n e^2}{\omega m} + \frac{\rho_{1\mathrm{F}} j_{a\q}^2}{ \omega} f\left(\frac{\omega - i /\tau}{2 \Delta_2}\right)
\end{equation}
with 
\begin{equation}
    f(z) = \frac{2 i \arcsin(z)}{z \sqrt{1 - z^2 }}
\end{equation}
it may be verified that in the limit of $\tau \to \infty$ (i.e. in the limit of $z \in \mathbb{R}$) this coincides with the previous result~\eqref{seq:sigma_clean}.

In the limit of large $\tau$, in the vicinity of the gap edge, this form simplifies
\begin{equation}
     \sigma_{aa}(\omega) \sim \frac{i n e^2}{2 \Delta_2 m} + \frac{\pi \rho_{2\mathrm{F}} j_{a\q}^2}{2 \Delta_2} \sqrt{\frac{\Delta_2}{\omega - 2 \Delta_2-i/\tau}}
\end{equation}
where $\sim$ indicates asymptotic equality when the limit of large $\tau$ is taken with $\tau(\omega - 2 \Delta_2)$ held fixed. This reveals the maximum height of $\sigma_{aa}(2\Delta_2) = O(\sqrt{\tau})$ which holds for the complex conductivity, and for the real and imaginary parts individually.

We comment that this approach does not recover the Case I response, i.e. the term which reproduces Drude conductivity in the limit $\Delta_2 \to 0$. This term must be obtained by other means~\cite{bardeen1957theory,schrieffer2018theory,tinkham2004introduction,mattis1958theory,abrikosov1959theory,nam1967theory,zimmermann1991optical,dressel2013electrodynamics,chen1992transport,chen1993theory,lee1985disordered}.

\section{Exact diagonalisation}
\label{app:ED}

\begin{figure*}
    \centering
    \includegraphics[width=\linewidth]{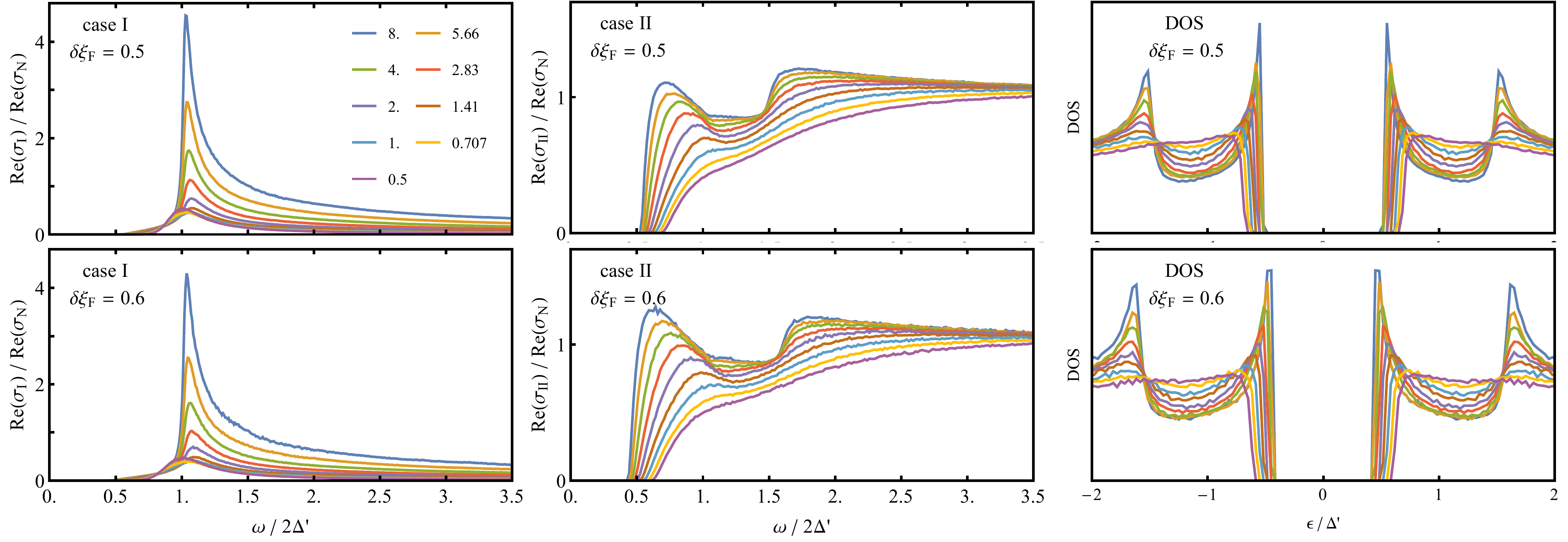}
    \caption{\emph{Additional data, as in Fig.~\ref{Fig:disorder}}: Here we show data, as in Fig.~\ref{Fig:disorder}, for additional parameters. Data is shown for two different values of the current, parameterised by $\delta \xi_{2\mathrm{F}} = q k_{2\mathrm{F}}/2m_2 = 0.5,0.6$ (top and bottom row respectively). The columns show the Case I part of the optical conductivity, the Case II part of the optical conductivity, and the density of states respectively. The Case I part is not significantly effected by vary the current but far a small change to the overall scale factor. Increasing the current does however change the indirect gap, moving the onset of the Case II part to lower frequencies (second column), whereas in terms of the quasiparticle density of states (third column) ones sees the inner pair of edges (at $\epsilon = \pm |\Delta - \delta \xi_{2\mathrm{F}}|$)  move to lower frequencies, and the outer pair (at $\epsilon = \pm |\Delta + \delta \xi_{2\mathrm{F}}|$) to higher frequencies.}
    \label{Fig:extra_disorder_data}
\end{figure*}

\begin{figure}[t!]
    \centering
    \includegraphics[width=0.9\linewidth]{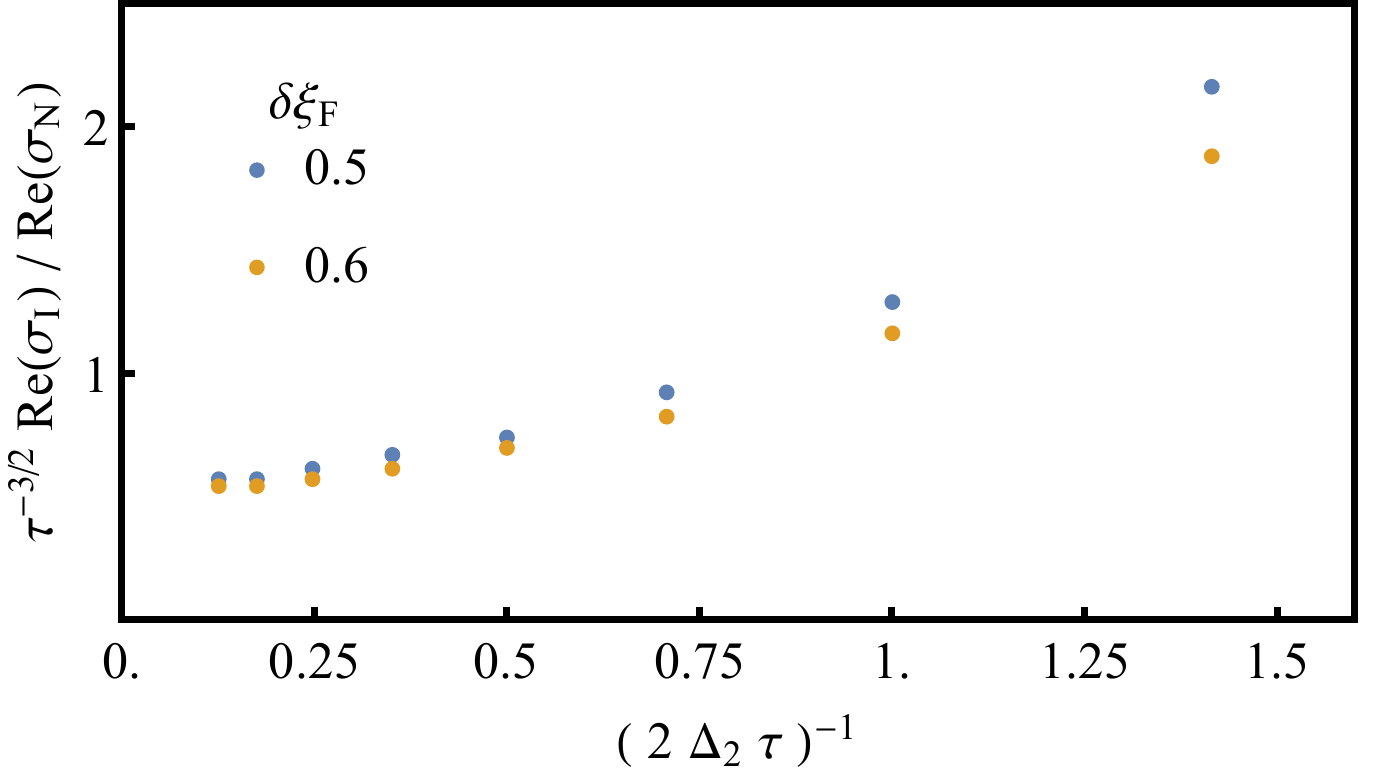}
    \caption{
    \emph{Scaling of the peak value of the Case I response}: Here we show the peak values of the Case I response (left column of Fig~\ref{Fig:extra_disorder_data}) re-scaled by $\tau^{-3/2}$. The two series correspond to distinct values of the current (values of $\delta \xi_{2\mathrm{F}}$ inset). The data displays apparent asymptotic convergence to a finite constant, confirming the relationship $\Re(\sigma_{\mathrm{I}}) / \Re(\sigma_\mathrm{N}) = O(\tau^{3/2})$ discussed in the main text, and derived in App.~\ref{app:sigma_deriv}. 
    }
    \label{Fig:scaling}
\end{figure}

In this section we provide details and additional data regarding the random matrix model used to numerically obtain forms for the dissipative part of the optical conductivity in Fig.~\ref{Fig:disorder} for a superconducting wire.

Data for the cases of $\delta \xi_{2\mathrm{F}} = q k_\mathrm{2F}/2m_2 = 0.5,0.6$ is provided in Fig.~\ref{Fig:extra_disorder_data}, while Fig.~\ref{Fig:scaling} provides numerical confirmation of the scaling of the current induced absorption peak stated in~\eqref{eq:scaling}.

\subsection{Numerical details}

We begin from the BdG Hamiltonian~\eqref{eq:h_bdg_int}. For simplicity we assume a hierarchy of scales that allows us to consider effects due to the second band only. Specifically we work in the frequency window $\omega \in [-\omega_{\mathrm{c}},\omega_{\mathrm{c}}]$ such that
\begin{equation}
    2 \Delta_2 + \frac{q k_\mathrm{2F}}{2 m_2} + \frac{1}{\tau} < \omega_{\mathrm{c}} < 2 \Delta_1  - \frac{q k_\mathrm{1F}}{2 m_1} - \frac{1}{\tau}.
\end{equation}
In words: we assume that $\omega_{\mathrm{c}}$ is at least $\tau^{-1}$ above the larger indirect gap of the second band, and at least $\tau^{-1}$ below the smaller indirect gap of the first band. In this frequency range the current operator for the first band has no spectral weight, and we may calculate the optical conductivity considering consider only the current operator for the second band. At the end of this section we discuss violations of this assumption.

We thus consider only the Hamiltonian of the second band~\eqref{eq:h_bdg_int}. In the present context, where momentum is not conserved, it is convenient to write this as a block matrix
\begin{equation}
    \tH = \Psi^\dagger \tilde{h} \Psi.
\end{equation}
where $\Psi$ is the vector of all $\psi_\k$. It is convenient to choose a basis in which $\tilde{h}$ takes the block diagonal form
\begin{equation}
    \tilde{h} = \begin{pmatrix}
    \tilde{h}_0 + \z \delta \xi_{2\mathrm{F}}  & \tilde{\Delta}
    \\
    \tilde{\Delta}^\dagger & -\tilde{h}_0^T - \z \delta \xi_{2\mathrm{F}}
    \end{pmatrix}
\end{equation}
where here and throughout $\sigma_\alpha$ are the usual pauli matrices. Here $\tilde{h}$ is denoted as block matrix with the top left (bottom right) blocks correspond to the particle (hole) sector, and the off-diagonal blocks correspond to the pair creation/breaking processes. Within the particle sector $\tilde{h}_0$ there are two further block---left movers and right movers (which have a relative offset due to the Doppler shift $\z \delta \xi_{2\mathrm{F}}$), and within each of these blocks there are $d$ states labelled by their different crystal momenta $k$---thus $\tilde{h}$ isa $4 d \times 4 d$ matrix.

The normal phase Hamiltonian is the sum of kinetic and potential energy terms
\begin{equation}
    \tilde{h}_0 = T_0 + V_0
\end{equation}
In the one dimensional setting considered here, the kinetic energy $T_0$ may in turn be expressed as a $2\times 2$ block diagonal matrix, with the two blocks corresponding to left and right movers respectively
\begin{equation}
    \begin{aligned}
    T_0 & = t_0 \sigma_0, \qquad t_0 = v_{2\mathrm{F}} (\hat{k} - k_{2\mathrm{F}}) \sigma_0 
    \end{aligned}
\end{equation}
where $\hat{k}$ is a matrix which measures the \emph{unsigned} momentum. For exact diagonalisation we keep only a finite number of states in the vicinity of the Fermi-surface, such that $\hat{k}$ is a $d \times d$ diagonal matrix whose diagonal values (i.e. eigenvalues) $k_\nu$ are evenly spaced over the interval specified by $E_\nu = v_{2\mathrm{F}} (k_\nu - k_{2\mathrm{F}}) \in [-E_\mathrm{c},E_\mathrm{c}]$, where we use an energy cutoff $E_\mathrm{c} = \omega_\mathrm{c} + \Delta_2 + \delta \xi_{2\mathrm{F}} + \lambda_V$ where $\lambda_V = v\sqrt{2 d}$ is the disorder bandwidth. $E_\mathrm{c}$ is chosen to exceed $\omega_\mathrm{c}$ by an additional buffer

The pairing potential matrix may be written in this basis
\begin{equation}
    \tilde{\Delta} = \begin{pmatrix}
    0 & \Delta_2 \x
    \\
    \Delta_2 \x & 0
    \end{pmatrix} =   \Delta_2 \x \tx
\end{equation}
where $\Delta$ is the scalar paring potential.

The scattering potential $V_0$ is a $2d \times 2d$ matrix which mixes couples left and right moving electrons. As the scattering potential is assumed to be real in the position basis, the matrix $V$ satisfies
\begin{equation}
    \x V_0 \x = V_0^*
    \label{seq:xVx}
\end{equation}
where $\x$ is the usual Pauli matrix, in this context implementing coordinate inversion, sending left movers to right movers and vice versa. We use a random matrix model of the scattering potential, assuming it induce a statistically identical coupling between all low energy momentum modes. Specifically, we use 
\begin{equation}
    V = U R U^\dagger, U = \frac{1}{\sqrt{2}}\begin{pmatrix}
    1 & 1
    \\
    i & -i
    \end{pmatrix}
\end{equation}
where $R=R^T$, one may verify that $V$ satisfies~\eqref{seq:xVx} as required. We take $R$ to be a Gaussian orthogonal matrix, i.e. the matrix elements $R_{ij}$ are real, identically distributed Gaussian random numbers with correlations
\begin{equation}
    [R_{ij} R_{nm}] = v^2 \left( \delta_{in}\delta_{jm} + \delta_{im}\delta_{jn} \right).
\end{equation}
where $v$ sets the typical scale of the matrix elements of $V$. The matrix element scale $v$ is related to the scattering time $\tau$ by Fermi's Golden rule
\begin{equation}
    1 / \tau = 2 \pi \rho v^2
\end{equation}
where $\rho = d/E_\mathrm{c}$ is the density of states coupled by the disorder, i.e. of $T_0$. We note that for the Fermi Golden rule result to be valid, we must be in the regime where it has a non-perturbative effect on $T_0$, the kinetic term, $\rho v = d v/ E_\mathrm{c} \gg 1$. However we also want to remain in he regime where the disorder does not alter the density of states of the minimal model---that is the typical eigenvalue of $T_0$ is of scale $\lambda_T = E_\mathrm{c}/\sqrt{3}$, far exceeds the typical eigenvalue scale of $V_0$, $\lambda_V = v\sqrt{2d}$, yielding the condition $E_\mathrm{c} \gg  v \sqrt{6d}$.

The current operator is given by~\eqref{eq:Jmf_2band}. We are interested only in the current carried by the second band, for which, in the present notation, the BdG matrix may be decomposed into its case I and case II pieces as
\begin{equation}
\begin{aligned}
    j_{2} & = j_{\mathrm{I}} + j_{\mathrm{II}}
    \\
    j_{\mathrm{I}} & = \frac{e (\delta v_{2} - v_\mathrm{c} )}{V} \tz
    \\
    j_{\mathrm{II}} & = \frac{e v_\mathrm{2F}}{V} \tz \z
\end{aligned}
\end{equation}

Finally we calculate the optical conductivity. We divide the dissipative part of the optical conductivity into corresponding case I and case II terms
\begin{equation}
    \Re [\sigma(\omega)] = \Re [\sigma_{\mathrm{I}}(\omega)] + \Re [\sigma_{\mathrm{II}}(\omega)]
\end{equation}
Subsequently, we obtain Fig.~\ref{Fig:disorder} numerically by diagonalising $\tilde{h}$, to obtain the diagonal orbitals $\tilde{h} \ket{\epsilon_\nu} = \epsilon_\nu \ket{\epsilon_\nu}$ and use the relation
\begin{equation}
    \Re [\sigma_{m}(\omega)] = \pi V \sum_{ \epsilon_\mu <0< \epsilon_\nu} |\bra{\epsilon_\mu} j_m \ket{\epsilon_\nu}|^2\delta(\omega - \omega_{\mu\nu})
\end{equation}
where $\omega_{\mu\nu} = \epsilon_\mu - \epsilon_\nu$, and $m \in \{ \mathrm{I},\mathrm{II} \}$, and the cross term disorder averages to zero (see App.~\ref{eq:Cross_term})
\begin{equation}
    \begin{aligned}
    \overline{\Re [\sigma_{\mathrm{cross}}(\omega)]} & = \pi V \overline{ \sum_{ \epsilon_\mu <0< \epsilon_\nu} \bra{\epsilon_\mu} j_{\mathrm{I}} \ket{\epsilon_\nu} \bra{\epsilon_\nu} j_{\mathrm{II}} \ket{\epsilon_\mu} \delta(\omega - \omega_{\mu\nu})}
    \\
    & = 0.
    \end{aligned}
\end{equation}
Repeating this calculation for $\Delta = 0$, $\Jsc = 0$ provides $\sigma_\mathrm{N}$ as required.

Finally we discuss parameters. As noted in previous paragraphs, the parameter regimes where this model is accurate are constrained. Specifically we require
\begin{enumerate}
    \item The disorder strength is non perturbative $\rho v = d v / E_\mathrm{c} \gg 1$.
    \item The disorder bandwidth is small compared to the window of kinetic energies $\lambda_\mathrm{T}/\lambda_\mathrm{V} = E_\mathrm{c} /  (v \sqrt{6 d}) \gg 1$. 
    \item We need sufficient density of states to observed the expulsion of levels from the superconducting gap $2 \Delta_2 \rho  = 2 d \Delta_2/E_\mathrm{c} \gg 1$.
    \item The energetic bias determining the current must also be non perturbative, requiring $\sqrt{d} \delta \xi_{2\mathrm{F}}/E_\mathrm{c} \gg 1$. This is the condition that the matrix elements of $\delta \xi_{2\mathrm{F}} \z \tz$ (which encode the energy bias setting the finite current) are of scale $\delta \xi_{2\mathrm{F}}/ \sqrt{d}$ far exceed the density of states in the electron/hole sectors $\rho = d / E_\mathrm{c}$.
\end{enumerate}
When these constraints are satisfied, the calculation of the optical conductivity is accurate for $\omega < \omega_\mathrm{c}$. For the numerics presented in the main text, we use these constraints, determining $x$ to satisfy $x \gg 1$ if $x>2$. Specifically we set $\Delta = 1$, $E_\mathrm{c} = 8.5, 8.6$, and $d = 4624, 3450$ for the data with $\delta \xi_{2\mathrm{F}} = 0.5, 0.6$ respectively.
\newline

\subsection{Cancellation of the cross term}
\label{eq:Cross_term}

Sufficient conditions for the case I/II cross term to average to zero is 
\begin{enumerate}
    \item The kinetic energy $\xi_\k$ may be linearised at the Fermi surface $\xi_\k = \v_{2\mathrm{F}} (\k - \k_{2\mathrm{F}})$
    \item The scattering matrix elements are identically distributed, and symmetric about $0$ (i.e. $V$ and $-V$ occur with equal probability).
\end{enumerate}
To see this, as before let $T_0 = t_0 \sigma_0$, $V$ denote the kinetic and disorder terms in $\tH$. Let also $u$ denote the normal involution ($u u^\dagger = u^\dagger u = u^2 = (u^\dagger)^2 = 1$) which changes the sign of the kinetic energy $u t_0 u = - t_0$. One may show that two Hamiltonians $\tH$ and $\tH'$ which are identical but for their disorder realisations $V$ and $V' = - u V u$ make exactly cancelling contributions to the case I/II cross term
\begin{equation}
    \Re [\sigma_{\mathrm{cross}}(\omega)] = \pi V \!\!\sum_{ \epsilon_\mu <0< \epsilon_\nu} \bra{\epsilon_\mu} j_\mathrm{I} \ket{\epsilon_\nu} \bra{\epsilon_\nu} j_\mathrm{II} \ket{\epsilon_\mu} \delta(\omega - \omega_{\mu\nu})
\end{equation}
Thus if $V$ and $V'= - u V u$ occur in the disorder ensemble with equal probabilities the cross term averages to zero. The first condition above is sufficient for the involution $u$ to exist, and the second is sufficient for $V$ and $V' = - u V u$ to occur with equal measure in the disorder ensemble.

\end{document}